# Marcus rate for electron transfer and the Goldilocks principle

Lev Mourokh[1] and Seth Lloyd[2]

[1]*Department of Physics, Queens College of the City University of New York, Flushing, NY 11367, USA.*

[2]*Department of Mechanical Engineering and Research Lab for Electronics, Massachusetts Institute of Technology, Cambridge, MA 02139, USA.*

We examine electron transfer between two quantum states in the presence of a dissipative environment represented as a set of independent harmonic oscillators. For this simple model, the Marcus transfer rates can be *derived* and we show that these rates are associated to an explicit expression for the environment correlation time. We demonstrate that as a manifestation of the Goldilocks principle, the optimal transfer is governed by a single parameter which is equal to just the inverse root square of two.

The conception of an electron transfer lies at the heart of many biological processes, chemical reactions, and electronic device operations. To explain the rates of chemical reactions, Rudolf Marcus developed an original theory of the electron transfer [1] which eventually brought him the Nobel Prize. These rates were obtained from the geometrical representation using the Franck-Condon principle and can be applied to numerous systems [2].

The Goldilocks principle proposed in Ref. [3-4] declares that biological systems are driven by natural selection to the conditions where the interaction with the environment is "just right" to attain maximum transport efficiency. This principle can be applied both to quantum and classical systems. In the former case, the interaction with the environment leads to decoherence, while in the later situation dissipation and fluctuations are induced. Numerical simulations of excitonic transport in the Fenna-Matthews-Olson photosynthetic complex (FMO) [4-7] show that transport rates attain a broad maximum as a function of the strength of environmental noise. In [3-4] it was shown both numerically [3] and by a general quantitative theory [4] that the rate of quantum transport is governed by a single parameter,

$$\Lambda = \frac{\lambda T}{\gamma_c \Delta\varepsilon}. \tag{1}$$

Here, $T$ is the temperature, $\lambda$ is the reorganization energy of the environment, $\gamma_c = 1/\tau_c$ is its inverse correlation time, and $\Delta\varepsilon$ is the characteristic energy separation scale. Optimal transport occurs when the Goldilocks parameter is of the order of one. In addition, the general theory [4] predicts that in highly decohering environments, $\Lambda \gg 1$, the rate of transition between neighboring sites should drop as $1/\Lambda$.

In the present paper, we discuss a simple model of the electron transfer between two states coupled to the environment in the form of a set of independent harmonic oscillators. We show that the Marcus rates can be obtained for this system from the equations of motion for the electron operators with the approximations of the weak coupling of the electron states and slow environment dynamics. The environment correlation time appears naturally from the microscopic consideration, so all the parameters involved in Eq. (1) can be determined at the point of the optimal transfer. Correspondingly, the Goldilocks parameter can be calculated and it is equal exactly to the inverse root square of two. Moreover, the high-temperature transition rate goes as $1/\sqrt{\lambda T} \sim 1/\Lambda$, confirming the predictions of the general theory of transport in [4].

The Hamiltonian of the systems under interest is given by

$$H = E_1 a_1^+ a_1 + E_2 a_2^+ a_2 - \Delta a_1^+ a_2 - \Delta^* a_2^+ a_1 + \sum_j \frac{p_j^2}{2m_j} + \sum_j \frac{m_j \omega_j^2}{2}\left(x_j - C_{1j} a_1^+ a_1 - C_{2j} a_2^+ a_2\right)^2, \quad (2)$$

where $a_\sigma^+ / a_\sigma$ are the electron creation/annihilation operators for the σ-state (σ = 1,2), $E_\sigma$ are the energies of these states, $\Delta$ is the transfer amplitude, $p_j$ and $x_j$ are the momentum and coordinate of the harmonic oscillator with the mass $m_j$ and the frequency $\omega_j$, and $C_{\sigma j}$ are the coupling strengths. After the unitary transformation

$$H' = U^+ H U; \quad U = \exp\left\{-i \sum_j p_j \left(C_{1j} a_1^+ a_1 + C_{2j} a_2^+ a_2\right)\right\}, \quad (3)$$

the Hamiltonian has the form

$$H = E_1 a_1^+ a_1 + E_2 a_2^+ a_2 - \Delta e^{i\xi} a_1^+ a_2 - \Delta^* e^{-i\xi} a_2^+ a_1 + \sum_j \left(\frac{p_j^2}{2m_j} + \frac{m_j \omega_j^2 x_j^2}{2}\right), \quad (4)$$

with the stochastic phase

$$\xi = \sum_j p_j \left(C_{1j} - C_{2j}\right). \quad (5)$$

Equations of motion derived from the Hamiltonian, Eq. (4), are given by

$$i\dot{a}_1 = E_1 a_1 - \Delta e^{i\xi} a_2,$$
$$-i\dot{a}_1^+ = E_1 a_1^+ - \Delta^* e^{-i\xi} a_2^+,$$
$$i\dot{a}_2 = E_2 a_2 - \Delta^* e^{-i\xi} a_2, \tag{6}$$
$$-i\dot{a}_2^+ = E_2 a_2^+ - \Delta e^{i\xi} a_1^+,$$

with the formal solutions

$$a_1(t) = a_1^{(0)}(t) - \Delta \int_{-\infty}^{t} dt_1 G_1^r(t,t_1) e^{i\xi(t_1)} a_2(t_1),$$
$$a_1^+(t) = a_1^{+(0)}(t) - \Delta^* \int_{-\infty}^{t} dt_1 G_1^a(t,t_1) e^{-i\xi(t_1)} a_2^+(t_1),$$
$$a_2(t) = a_2^{(0)}(t) - \Delta^* \int_{-\infty}^{t} dt_1 G_2^r(t,t_1) e^{-i\xi(t_1)} a_1(t_1), \tag{7}$$
$$a_2^+(t) = a_2^{+(0)}(t) - \Delta \int_{-\infty}^{t} dt_1 G_2^a(t,t_1) e^{i\xi(t_1)} a_1^+(t_1),$$

where free operators $a_\sigma^{+(0)}(t)/a_\sigma^{(0)}(t)$ describe the time evolution without transfer to another state, $G_\sigma^r(t,t_1) = \langle -i[a_\sigma^{(0)}(t), a_\sigma^{+(0)}(t_1)]_+ \rangle$ and $G_\sigma^a(t,t_1) = \langle i[a_\sigma^{+(0)}(t), a_\sigma^{(0)}(t_1)]_+ \rangle$ are the retarded and advanced Green's functions, respectively, and $[...,...]_+$ is the anticommutator. Here, the angular brackets mean both the quantum-mechanical and thermal averaging procedures.

The time evolution of the averaged population of the first state,

$$\langle \dot{n}_1 \rangle = \langle \dot{a}_1^+ a_1 + a_1^+ \dot{a}_1 \rangle = i\Delta \langle e^{i\xi} a_1^+ a_2 \rangle - i\Delta^* \langle e^{-i\xi} a_2^+ a_1 \rangle, \tag{8}$$

can be evaluated using the formula

$$i\Delta \langle e^{i\xi(t)} a_1^+(t) a_2^{(0)}(t) \rangle = i\Delta \int_{-\infty}^{\infty} dt_1 \langle a_2^{+(0)}(t_1) a_2^{(0)}(t) \rangle \left\langle \frac{\delta(e^{i\xi(t)} a_1^+(t))}{\delta a_2^{+(0)}(t_1)} \right\rangle, \tag{9}$$

where the functional derivative can be expressed as a commutator [8],

$$\left\langle \frac{\delta(e^{i\xi(t)} a_1^+(t))}{\delta a_2^{+(0)}(t_1)} \right\rangle = i \langle [e^{i\xi(t)} a_1^+(t), \Delta^* e^{-i\xi(t_1)} a_1(t_1)]_- \rangle \theta(t - t_1), \tag{10}$$

with $\theta(t - t_1)$ being the unit step function. Correspondingly, we obtain

$$\langle \dot{n}_1 \rangle = |\Delta|^2 \int_{-\infty}^{t} dt_1 \Big( \langle a_2^{+(0)}(t_1) a_2^{(0)}(t) \rangle \langle a_1(t_1) a_1^+(t) \rangle \langle e^{-i\xi(t_1)} e^{i\xi(t)} \rangle$$
$$+ \langle a_2^{(0)}(t_1) a_2^{+(0)}(t) \rangle \langle a_1^+(t_1) a_1(t) \rangle \langle e^{i\xi(t_1)} e^{-i\xi(t)} \rangle$$
$$- \langle a_2^{(0)}(t) a_2^{+(0)}(t_1) \rangle \langle a_1^+(t) a_1(t_1) \rangle \langle e^{i\xi(t)} e^{-i\xi(t_1)} \rangle \tag{11}$$
$$- \langle a_2^{+(0)}(t) a_2^{(0)}(t_1) \rangle \langle a_1(t) a_1^+(t_1) \rangle \langle e^{-i\xi(t)} e^{i\xi(t_1)} \rangle \Big).$$

For the case of weak transfer coupling, the correlators of full electron operators in Eq. (11) can be replaced by that of free operators, which can be evaluated as

$$\langle a_\sigma^{+(0)}(t_1)a_\sigma^{(0)}(t)\rangle = \exp\{-iE_\sigma(t-t_1)\}\langle a_\sigma^{+(0)}(t)a_\sigma^{(0)}(t)\rangle = \exp\{-iE_\sigma(t-t_1)\}\langle n_\sigma(t)\rangle,$$
$$\langle a_\sigma^{(0)}(t_1)a_\sigma^{+(0)}(t)\rangle = \exp\{iE_\sigma(t-t_1)\}\langle a_\sigma^{(0)}(t)a_\sigma^{+(0)}(t)\rangle = \exp\{iE_\sigma(t-t_1)\}(1-\langle n_\sigma(t)\rangle).$$
(12)

To determine the correlator of phase factors, we use the Baker-Hausdorf formula

$$\exp\{-i\xi(t)\}\exp\{i\xi(t_1)\} = \exp\{-i(\xi(t)-\xi(t_1))\}\exp\left\{\frac{1}{2}[\xi(t),\xi(t_1)]_-\right\},$$
(13)

where the commutator

$$\frac{1}{2}[\xi(t),\xi(t_1)]_- = -i\sum_j m_j\omega_j(C_{1j}-C_{2j})^2\sin\omega_j\tau$$
(14)

is determined using the free-evolving oscillator operators,

$$p_j(t) = p_j(t_1)\cos\omega_j\tau - m_j\omega_j x_j(t_1)\sin\omega_j\tau,$$
(15)

where $\tau = t - t_1$. For the Gaussian statistics of the system of independent oscillators, the characteristic functional has the form

$$\langle\exp\{-i(\xi(t)-\xi(t_1))\}\rangle = \exp\left\{-\langle\xi^2\rangle + \frac{1}{2}\langle[\xi(t),\xi(t_1)]_+\rangle\right\},$$
(16)

with

$$\frac{1}{2}\langle[\xi(t),\xi(t_1)]_+\rangle = \frac{1}{2}\sum_j(C_{1j}-C_{2j})^2\langle[p_j(t),p_j(t_1)]_+\rangle = \sum_j\langle p_j^2\rangle(C_{1j}-C_{2j})^2\cos\omega_j\tau.$$
(17)

The variance of the momentum of the $j$-th oscillator is given by

$$\langle p_j^2\rangle = \frac{m_j\omega_j}{2}\coth\left(\frac{\omega_j}{2T}\right).$$
(18)

Introducing the reorganization energy associated with the electron transfer, as

$$\lambda = \sum_j \frac{m_j\omega_j^2}{2}(C_{1j}-C_{2j})^2,$$
(19)

and assuming slow fluctuations of the environment ($\omega_j \ll T$, $\omega_j\tau_c \ll 1$, where $\tau_c$ is the bath correlation time), so $\sin\omega_j\tau = \omega_j\tau$, $1-\cos\omega_j\tau = \omega_j^2\tau^2/2$, and $\coth(\omega_j/2T) = 2T/\omega_j$, we finally obtain

$$\langle\exp\{-i\xi(t)\}\exp\{i\xi(t_1)\}\rangle = \exp\{-i\lambda\tau\}\exp\{-\lambda T\tau^2\}. \tag{20}$$

It is evident from Eq. (20) that the bath correlation time is $\tau_c = 1/\sqrt{\lambda T}$. Substituting Eqs. (13,20) into Eq. (12) and integrating with respect to $\tau$, we obtain the rate equation in the self-consistent form, as

$$\begin{aligned}\langle\dot{n}_1\rangle &= \kappa(E_2 - E_1 + \lambda)\langle n_2\rangle(1-\langle n_1\rangle) - \kappa(E_1 - E_2 + \lambda)\langle n_1\rangle(1-\langle n_2\rangle) \\ \langle\dot{n}_2\rangle &= -\langle\dot{n}_1\rangle\end{aligned}, \tag{21}$$

where

$$\kappa(\Delta\varepsilon) = |\Delta|^2 \sqrt{\frac{\pi}{\lambda T}} \exp\left\{-\frac{(\Delta\varepsilon)^2}{4\lambda T}\right\} \tag{22}$$

is the well-known Marcus transfer rate [1]. It is evident from Eq. (21) that, as expected, the probability of the transfer is proportional to the occupation of the initial state and the likelihood of the final state to be empty with the Marcus rates being the proportionality coefficients. The total energy change including that of the reorganization of the environment is involved in the numerator of the exponent argument in Eq. (22). Note that for large $\lambda T$ the rate is proportional to $1/\sqrt{\lambda T} \sim 1/\Lambda$, confirming the prediction of [4].

To determine the Goldilocks parameter at the point of the optimal performance, we can take the derivative of the Marcus rate, Eq. (22), with respect to temperature and equalize it to zero. Correspondingly, $\Delta\varepsilon = \sqrt{2\lambda T}$. Inserting this expression and that for the bath correlation time into Eq. (1), we obtain

$$\Lambda = \frac{\lambda T}{\sqrt{\lambda T}\sqrt{2\lambda T}} = \frac{1}{\sqrt{2}}. \tag{23}$$

Accordingly, for *all* transfer events described by the Marcus rate, the temperature for the optimal performance can be determined from the Goldilocks parameter being just the inverse root square of two. In the set-up of the present paper, the electron transfer picture can be applied to the outer-sphere electrons for chemical reactions, electron transport through the chain of semiconductor quantum dots, HOMO-LUMO electron transfer in complex molecules, and so on. Moreover, similar approach can be applied to proton transport in proton pumps [9-11] or to exciton transport in photosynthetic complexes [12].

In conclusion, we considered a simple model for the electron transfer between two states in the presence of the environment in the form of the set of independent harmonic oscillators. From equations of motion for the electron operators averaged over the environment, we obtained that the transfer amplitude is given by the well-known Marcus transfer rate and determined the bath correlation time associated with this rate. We showed that the Goldilocks parameter at the point of the optimal performance equals to just the inverse root square of two, and that transition rates for $\Lambda \gg 1$ go as $1/\Lambda$, as predicted in [4]. We argue that these properties remain the same for all transfer events which can be described by the Marcus rate, so it has a broad applicability to numerous biological and chemical processes, as well as to the processes in electronic devices.

**Acknowledgments**

L. M. was partially supported by PSC-CUNY award 65245-00 43. S. L. was supported by DARPA and by Eni under the MIT Energy Initiative.